\documentstyle[aps,twocolumn,psfig]{revtex}

\begin{document}
\title{Influence of nanoscale regions on Raman spectra of complex perovskites}
\author{E.A. Rogacheva}
\address{A.F. Ioffe Physical Technical Institute, 194021 St.~Petersburg, Russia\\
Electronic mail: trirog@mail.wplus.net}
\date{\today}
\maketitle

\begin{abstract}
New approach on interpretation and processing of Raman spectra of complex
perovskites is suggested. Raman spectra of $PbMg_{1/3}Nb_{2/3}O_{3}$ and $%
PbSc_{1/2}Ta_{1/2}O_{3}$ are successfully described on the basis of the
phonon-confinement model and validity of the method is demonstrated on good
agreement of experimentally obtained and computer simulated spectra of
another complex perovskite, $BaMg_{1/3}Ta_{2/3}O_{3}$. The study showed that
the size of nanoscale regions is a kind of an inborn characteristic of the
sample and is not temperature dependent. For $PbSc_{1/2}Ta_{1/2}O_{3}$, two
special temperature points, approximately 400 K and 600 K are found from the
analysis of the temperature behaviour of obtained mode parameters.

\end{abstract}
\section{Introduction}

\noindent Complex perovskites form vast and promising group of compounds for
microelectronics, yet many features they show are still a battlefield for
scientists. Among the compounds with a well-known general formula $%
AB_{x}^{\prime }B_{1-x}^{\prime \prime }O_{3}$ one can find a spectrum of
cases, from relaxor ferroelectrics (or ferroelectrics with diffuse phase
transition), such as $PbMg_{1/3}Nb_{2/3}O_{3}$ ({\it PMN}) and $%
PbSc_{1/2}Ta_{1/2}O_{3}$ ({\it PST}) to antiferroelectrics ($%
PbMg_{1/2}W_{1/2}O_{3}$ ({\it PMW}) for example) \cite{smolensky}.
Historically, {\it PMN} and {\it PST} are most widely studied, due to their
relaxor behaviour. Numerous structural studies of such compounds were
performed by different techniques (x-ray scattering (see, for example, \cite
{prouzet}), HRTEM, dark field images \cite{bursill,reaney,harmer}) together
with neutron scattering \cite{glazunov}, synchrotron studies \cite
{vakhrushev}, and NMR studies \cite{glinchuk}, but still there is no
unanimous point of view on ''what makes a relaxor to be a relaxor''. In some
sense, the results of Raman studies make the problem even more confusing.

Every Raman study of a dielectric crystal leads to a question: to what space
group does this compound belong? To make a line assignment one needs at
least to determine a ''basis'' for that procedure, i.e. to deconvolute the
spectrum (and to prove whether this spectrum as a whole or a given part of the
spectrum is of first order). In some cases this part of the task is not
challenging, but for relaxors. As a matter of fact, in general, complex
perovskites' unit sell belongs to $Pm3m$ space group for which no first
order Raman scattering is allowed by group theory analysis \cite{lines}. But
it is well known that these compounds do exhibit first order Raman spectra 
\cite{bismayer,setter,sinyb,sinys,slkr}. Several attempts to make line
assignment were maid, and the most common point of view is that they are
consistent with $Fm3m$ space group \cite{slkr}. Such studies were performed
both on raw \cite{bismayer,setter,sinyb,sinys,slkr} and reduced Raman
spectra \cite{marssi} (although due to a presence of a very low-frequency
line raw spectra are not so informative), but these studies suffered from
one and the same problem, namely, the problem of extra lines. Different ways
to overcome this difficulty were suggested, and such lines were treated
either as arising due to distortions, or as second order regions of spectra.
Another surpassing detail for relaxors is that the lines are unusually broad
(40 cm$^{-1}$ as a FWHM is a normal value for relaxors) and posses a complex
structure of lines, that seems to be nearly a characteristic feature of
Raman spectra of the compounds understudy. Also some of Raman lines are
obviously asymmetric.

Besides, extensive study of temperature behaviour of Raman spectra of {\it %
PMN} and {\it PST} crystals showed that the spectra are temperature
dependent and anomalies correlated with other phenomena (such as specific
temperature behaviour of the acoustic phonon damping revealed by Brillouin
scattering studies and sensitivity of {\it PMN} spectra for anomalies in
complex dielectric response) were observed \cite{slkr}.

The present work is aimed on suggestion of a new approach to Raman spectra
of $AB_{x}^{\prime }B_{1-x}^{\prime \prime }O_{3}$ complex perovskites.

\section{Experimental}

Diagonal and parallel Raman spectra of {\it PMN} ({\it X(ZZ)Y} and {\it %
X(ZX)Y}) and {\it PST} ({\it Z(XX)Y} and {\it Z(XY)Y}) crystals were
measured with {\it X}, {\it Y} and {\it Z} being along the fourfold cubic
axes. {\it PST} crystals were obtained for two levels of ordering.
Investigations were carried out in a wide temperature range for both
compounds. The details of sample preparation and experiment setup can be
found elsewhere \cite{slkr}.

\section{Raman spectra of PMN and PST crystals}

Examples of Stokes parts of polarized Raman spectra of{\it \ PMN} and
disordered {\it PST} are presented in Fig. 1. In this paper only reduced
Raman spectra are discussed, so the word ''spectrum'' is actually for
''reduced spectrum'' unless specially stated. The reason is that dealing
with complex perovskites one inevitably has to deal with low frequency
region where Bose-Einstein population factor can not be neglected. Another
peculiarity to be mentioned beforehand is that, being well acquainted to the
results of group theory analysis and mode assignment performed for these
compounds, the author deliberately tries to avoid using appropriate mode
identifications (such as $F_{2g}$, $E_{g}$ and $A_{1g}$) when speaking of
some part of a spectrum. This may look rather clumsy through the text but is
believed to be necessary for a presentation of the approach.

One can easily observe some common features of these spectra. (Mind that at
495 K{\it \ PMN} is in ferrophase and at 373 K a disordered {\it PST} is in
paraphase.) The striking detail is that the likely regions of spectra have
practically the same frequency shifts. For example, apart from such obvious
lines as those at approximately 50 cm$^{-1}$ and 800 cm$^{-1}$, the bands in
the regions of 100 - 300 cm$^{-1}$ and 500 - 600 cm$^{-1}$ are more or less
likely. The lines at approximately 380 and 550 cm$^{-1}$ in {\it Z(YZ)X}
polarization are clearly resolved only for {\it PST}. According to
dielectric response data, disordered {\it PST} is a relaxor as well as {\it %
PMN}. Temperature evolution at least of the low frequency region of {\it PMN}
was already reported, so this time let us consider the temperature evolution
of {\it Z(YY)X} spectra of {\it PST} (Fig. 2). Quite obviously the spectrum
at 292 K differs from one at 741 K, and also a specific temperature
evolution is seen. As the population factor is already taken into account,
one can suppose that the whole spectra are the first order Raman spectra
since no band performs a drastically different temperature behaviour.

So, there are first order Raman spectra exhibiting unusually broad and
asymmetric lines and specific temperature evolution obtained from compounds
for which no first order Raman scattering is allowed.

\section{Interpretation of Raman spectra}

The considerations stated above already lead to conclusion that there is a
factor, which violates the Raman selection rules thus allowing the existence
of forbidden spectra. It means that for some reason the scattering observed
originates not only from the center of the Brillouin zone but from other
points of the Brillouin zone as well. As additional evidences for violation of
selection rules one can recall that the IR spectra \cite{setter} of{\it \ PMN%
} (as well as neutron diffraction studies \cite{lushnikov}) revealed
practically the same set of modes as for Raman spectra. Another fact, which
is worth mentioning, is that structural studies performed for {\it PST}
showed that there are significant deviations from $Fm3m$ symmetry \cite
{kishi}. And nevertheless these compounds exhibit practically identical
spectra. Many studies revealed the existence of nanoregions in {\it PMN} and 
{\it PST} \cite{boulesteixh,julin,cai}. Although the nature of such regions
is still under discussion and quite contradictory interpretations are
suggested \cite{egami}, nowadays the presence of nanoscale structure in
relaxors can not be denied. Recently on the basis of similarity of Raman
spectra of $BaMg_{1/3}Ta_{2/3}O_{3}$ ({\it BMT}) crystals and ceramics it 
was suggested that the Raman
scattering in {\it BMT} is determined by the short-range order in the
nanoscale microstructures \cite{bmt}.

The smallness of nanoregions can be, in principle, the factor, which
violates Raman selection rules. Indeed, in ideal infinite crystal only
phonons at the center of the Brillouin zone (with $\vec{q}\approx 0$) can be
observed due to crystal momentum conservation. In an imperfect crystal (and
there are no reasons to treat a complex perovskite as an ideal crystal)
phonons can be confined in space. Thus, according to the uncertainty
relation, there appears an uncertainty in the phonon momentum and phonons
with $\vec{q}>0$ can contribute to Raman signal. Provided that the region of
scattering is very small, scattering from practically the entire Brillouin
zone is allowed to contribute to the signal. To describe a frequency shift
and broadening of lines in Raman spectra in the case of scattering from
sufficiently small regions a model was developed nowadays known as spatial
correlation or phonon-confinement model \cite{ager,campbell,richter}. This
model is outlined briefly below.

The wave function of a phonon with a wave vector in an infinite perfect
crystal is
\begin{equation}
\phi \left( \vec{q}_{0},\vec{r}\right) =u\left( \vec{q}_{0},r\right) \exp
\left( -i\vec{q}_{0}\cdot \vec{r}\right) ,
\label{one}
\end{equation}
where $u\left( \vec{q}_{0},\vec{r}\right) $ is the periodicity of the
lattice. Let us suppose that the phonon is confined to the sphere of
diameter $L$. Such confinement can be accounted for by writing another wave
function $\psi $ instead of $\phi $%
\begin{equation}
\psi \left( \vec{q}_{0},\vec{r}\right) =A\exp \left\{ \frac{r^{2}}{2}/\left( 
\frac{L}{2}\right) ^{2}\right\} \phi \left( \vec{q}_{0},\vec{r}\right) =\psi
^{\prime }\left( \vec{q}_{0},\vec{r}\right) \cdot u\left( \vec{q}_{0},\vec{r}%
\right) , 
\label{two}
\end{equation}

where 
\begin{equation}
\left| \psi \right| ^{2}=A^{2}\exp \left\{ -r^{2}/\left( \frac{L}{2}\right)
^{2}\right\} . 
\label{three}
\end{equation}

That means that $\psi $ is confined to $\left| r\right| \leq L$ in the form
of Gauss distribution of width $\sqrt{\ln 2L}.And \psi ^{\prime }$ might be
expanded in a Fourier series: 
\begin{equation}
\psi \left( \vec{q}_{0},\vec{r}\right) =\int d^{3}qC\left( \vec{q}_{0},\vec{q%
}\right) \exp \left( i\vec{q}\vec{r}\right) , 
\label{four}
\end{equation}

where the Fourier coefficients are given by 
\begin{equation}
C\left( \vec{q}_{0},\vec{q}\right) =\frac{1}{(2\pi )^{3}}\int d^{3}r\psi
^{\prime }\left( \vec{q}_{0},\vec{r}\right) \exp \left( -i\vec{q}\vec{r}%
\right) . 
\label{five}
\end{equation}

Substitution of $\psi ^{\prime }$ from (2) to into (5) yields 
\begin{equation}
C\left( \vec{q}_{0},\vec{q}\right) =\frac{AL}{\left( 2\pi \right) ^{3/2}}%
\exp \left\{ -\frac{1}{2}\left( \frac{L}{2}\right) ^{2}\left( \vec{q}-\vec{q}%
_{0}\right) ^{2}\right\} . 
\label{six}
\end{equation}

Thus the $\psi ^{\prime }$ and $\psi $ seize to be Eigenfunctions of a
phonon wave vector $\vec{q}_{0}$, and become a superposition of
Eigenfunctions with $\vec{q}$ vectors in an interval $\left| q-q_{0}\right|
\leq \frac{1}{2L}$ centered at $\vec{q}_{0}$. For the chosen form of
confinement (3) the Eigenfunctions are weighted through the coefficients $%
C\left( \vec{q}_{0},\vec{q}\right) $, according to a Gauss distribution.

Thus, it is supposed that for the phonon transition matrix elements $\left|
\left\langle \vec{q}_{0}\left| \widehat{o}\right| \vec{q}\right\rangle
\right| ^{2}$ there appear non-vanishing values also for $\vec{q}\neq \vec{q}%
_{0}$ according to: 
\begin{equation}
\left| \left\langle \vec{q}_{0}\left| \widehat{o}\right| \vec{q}%
\right\rangle \right| ^{2}=\left| \left\langle \vec{q}_{0}\left| \widehat{o}%
\right| \vec{q}_{0}\right\rangle \right| ^{2}\cdot C\left( \vec{q}_{0},\vec{q%
}\right) , 
\label{seven}
\end{equation}

where $\left| \widehat{o}\right| $ is the phonon-phonon interaction
operator. To simplify the equation, when writing (7) it was assumed that $%
u\left( \vec{q},\vec{r}\right) =u\left( \vec{q}_{0},\vec{r}\right) $. Thus
the phonon confinement leads to relaxation of $\Delta q=0$ selection rule.

Normally when speaking of Raman spectra we actually deal with the excited optical phonon
in the center of the Brillouin zone $\left( \vec{q}=0\right) $. In case of
allowed $\vec{q}\neq o$ transiotions contributions with $\omega $ determined
by the dispersion relations $\omega \left( \vec{q}\right) $ will add to the
Raman spectrum. Obviously, the additional transitions with $\vec{q}\neq 0$
lead to a broadening of a Raman line and its shift towards higher or lower
frequencies. The direction of such frequency shift depends on a dispersion
relation for each phonon brunch and will be examined in more details further
on.

It is possible to write an expression for a Raman line on the basis of the
above mentioned considerations.The wave function of the confined phonon can
be expressed via the wave functuion of a phonon in an ideal infinite crystal
as (here the notations of \cite{ager} are used): 
\begin{equation}
\psi \left( \vec{q}_{0},\vec{r}\right) =W\left( \vec{r},L\right) \phi \left( 
\vec{q}_{0},\vec{r}\right) =\psi ^{\prime }\left( \vec{q}_{0},\vec{r}\right)
u\left( \vec{q}_{0},\vec{r}\right) ,
\label{eight} 
\end{equation}

where $W\left( \vec{r},L\right) $ describes confinement. Then the Fourier
coefficients are 

\begin{equation}
C\left( \vec{q}_{0},\vec{q}\right) =\frac{1}{(2\pi )^{3}}\int d^{3}r\psi
^{\prime }\left( \vec{q}_{0},\vec{r}\right) \exp \left( -i\vec{q}\vec{r}%
\right) 
\label{nine}
\end{equation}

\[
=\frac{1}{(2\pi )^{3}}\int d^{3}rW\left( \vec{r},L\right) \exp
\left( -i\left( \vec{q}-\vec{q}_{0}\right) \cdot \vec{r}\right) . 
\]

Let us stress again that the confined phonon wave function is a
superposition of plane waves with wave vectors $\vec{q}$ centered at $\vec{q}%
_{0}$. The Raman lineshape appears to be constructed by superposition of
Lorentzian line shapes (with the linewidth of an imagined ideal crystalline
media for the compound understudy) centerd at\ $\omega \left( \vec{q}\right) 
$ and weighted by the wave-vector uncertainty caused by confinement: 
\begin{equation}
I\left( \omega \right) \cong \frac{\int_{o}^{1}\frac{d^{3}\vec{q}\left|
C\left( 0,\vec{q}\right) \right| ^{2}}{\left[ \omega -\omega \left( \vec{q}%
\right) \right] ^{2}-\left( \frac{\Gamma _{0}}{2}\right) ^{2}}}{%
\int_{0}^{1}d^{3}\vec{q}\left| C\left( 0,\vec{q}\right) \right| ^{2}}, 
\label{ten}
\end{equation}

where $\omega \left( \vec{q}\right) $ is the phonon-dispersion curve, and $%
\Gamma _{0}$ is the imagined linewidth of an ideal crystalline media for the
compound understudy. It is important to notice that, unlike crystals like $%
KTaO_{3}$, $K_{1-x}Li_{x}TaO_{3}$ in case of complex perovskites we have no
sample compound that can give us the exact value of $\Gamma _{0}$. $\vec{q}%
=0 $ corresponds to the scattering from the center of the Brillouin zone and
the integration is carried over the entire Brillouin zone. With $%
L\rightarrow \infty $, $C\left( 0,\vec{q}\right) =\delta \left( \vec{q}%
\right) $ and $I\left( \omega \right) $ is a Lorentzian centerd at $\omega
\left( 0\right) $ and with a linewidth of $\Gamma _{0}$. Different functions
can be chosen to describe the localization, but, according to \cite{campbell}%
, for the case of sampling a signal from a number of regions of localization
a Gaussian (provided that the amplitude is $\cong 0$ on a border of a
region) is the most successful. Thus we have 
\begin{equation}
W\left( \vec{r},L\right) =\exp \left[ \frac{-8\pi ^{2}r^{2}}{L^{2}}\right] ,
\label{eleven}
\end{equation}
\begin{equation}
\text{ \ }\left| C\left( 0,\vec{q}\right) \right| ^{2}\cong \exp \left[ 
\frac{-q^{2}L^{2}}{4}\right] , 
\label{twelve}
\end{equation}

where $q$ is given in units of $2\pi /a$, where $a$ is the lattice constant
(approximately 4 \AA\ for complex perovskites), and $L$ is given in units of $%
a$.

Mind that further on instead of arbitrary intensity $I\left( \omega \right) $
\ Raman suscptibility $\chi \left( \omega \right) $ will be considered
obtained as 
\begin{equation}
\chi \left( \omega \right) =\frac{I\left( \omega \right) }{F\left( \omega
,T\right) }, 
\label{thirteen}
\end{equation}

where $F\left( \omega ,T\right) =\left[ n\left( \omega ,T\right) +1\right] $
for Stokes part of spectra and $F\left( \omega ,T\right) =n\left( \omega
,T\right) $ for anti-Stokes part respectively, and 
\begin{equation}
n\left( \omega ,T\right) =\left[ \exp \left( \hbar \omega /kT\right) -1%
\right] ^{-1}, 
\label{fourteen}
\end{equation}

where $\omega $ is the Raman shift and $T$ is the temperature.

So, the Raman line shape for the case of a phoon confined to a sphere of
diameter $L$, and a Gaussian confinement is 
\begin{equation}
\chi \left( \omega \right) \cong \frac{\int_{0}^{1}\frac{dq\exp \left(
-q^{2}L^{2}/4\right) 4\pi q^{2}}{\left[ \omega -\omega \left( q\right) %
\right] ^{2}+\left( \Gamma _{0}/2\right) ^{2}}}{\int_{0}^{1}dq\exp \left(
-q^{2}L^{2}/4\right) 4\pi q^{2}}, 
\label{fifthteen}
\end{equation}

where $\omega \left( q\right) $ is an approximate one-dimensional
phonon-dispersion curve. Unfortunately, there are no reported experimental
data on phonon-dispersion curves in complex perovskites, but it can be
easily shown that for such calculations the quadratic term of polinomial
approximation is playing the key role and other terms can be neglected. So,
the dispersion curve for (14) can be written as 
\begin{equation}
\omega \left( q\right) =\omega _{0}\pm Q\omega ^{2}, 
\label{sixteen}
\end{equation}

where $Q$ is the quadratic term and the sign before $Q$ in fact determines
whether the line suffers a shift to low freuencies (minus) or to higher
frequencies (plus). Fig. 3 illustrates alteration of frequency position and
shape of a Raman line with different $L$ and $Q$.

\section{Results and discussion}

The initial aim of the present study was to describe the experimental Raman
spectra of {\it PMN} and {\it PST} with a fixed set of modes on the basis of
spatial-correlation model. An example of such calculation is presented in
Fig. 4. In the contribution only a region of 0 - 400 cm$^{-1}$ is
considered. The results of interpretation of the rest of the frequency
region (400 - 1000 cm$^{-1}$) will be reported later, but it is possible to
state that these results do not contradict to the suggested interpretation.
Besides, such sharing of spectra regions does not lead to mistakes in
calculations. Similar calculations were performed for all experimentally
obtained spectra. The first important result is that the size of nanoregions
is a kind of ''inborn'' characteristic of the sample (or, generally
speaking, of the type of compound) and is not temperature dependent. Thus,
for {\it PMN} and {\it PST} disordered $L=4$ and for {\it PST} ordered $L=5$. To check
the validity of such approximation in the results obtained for one of a
high-temperature spectra of {\it PST} was taken equal to 12 (that value
influences lineshapes not so dramatically) and both {\it Z(YY)X} and {\it %
Z(YZ)X} lines were calculated to make the simulated spectra effectively
unpolarized. The results appeared to be in good agreement with the spectra
experimentally obtained by authors of \cite{bmt} for non-polarized spectra
of crystalline {\it BMT} (Fig. 5).

One can readily notice that in the frequency range of 100 - 200 cm$^{-1}$
the simulated Raman lines are different from those in the frequency region
of 150 - 250 cm$^{-1}$ on the experimentally obtained spectrum, although
the number of lines is the same. It appeared that for disordered {\it PST}
there are modes in the region of 100 - 200 cm$^{-1}$, which demonstrate an
alteration of parameters (especially of the value of the line frequency
position) with temperature growth. One of these modes seems to be a
''softening'' mode, because its positions changes rapidly from 260 cm$^{-1}$
at 550 K to 140 cm$^{-1}$ at 600 K, while its intensity significantly
decreases in the range of 300 - 500 K. For the ordered PST\ also one mode
with similar behaviour was found. It changes its position from 160 cm$^{-1%
\text{ }}$at 300 K down to 90 cm$^{-1}$at 600 K.

The analysis of the obtained parameters of the phonon modes showed some
anomalies in their temperature behavior. The common feature of these
anomalies is that they all happen in the region of approximately 350 - 600 K
and the main temperature points appear to be in the vicinty of
400~K and 600~K. The point 400~K is in good agreement with suggested 
estimation of Curie-Weiss temperature
performed on the basis of experimental investigation of dielectric
permittivity of {\it PST} crystals \cite{vieland}. As for 600 K, anomalies
of the refractive index starting at approximately 600 K were reported by the
authors of \cite{korshunov}.

More detailed analysis of the temperature behaviour of mode parameters will
be published later.

\section*{Acknowledgement}
The work was supported by INTAS Project 96/167, INTAS Fellowship grant for
Young Scientists No YSF~98-75. The author is extremely thankful to the Royal
Institute of Technology for technical support and to Dr. E.Obraztsova and
Dr. A.Tagantsev for helpful and inspiring discussions.


\section*{Figure Captions}

\begin{figure}[h]
\caption{a) reduced {\it X(ZZ)Y} (solid line) and {\it X(ZX)Y} (dotted line)
Raman spectra of {\it PMN} at 495 K; b) reduced {\it Z(XX)Y} (solid line)
and {\it Z(XY)Y} (dotted line) spectra of the disordered {\it PST} at 373 K.}
\end{figure}

\begin{figure}
\caption{Temperature evolution of the reduced {\it Z(XX)Y} Raman spectra of
the disordered {\it PST}.}
\end{figure}

\begin{figure}
\caption{Alteration of a Raman line shape and position under relaxation of  $%
\vec{q}=0$ rule; $L$ is a nanoregion size (in units of $a$) to which a
phonon is confined and $Q$ is a quadratic term (eq. 16). Here a) and c)
correspond to cases of different $L$ and constant $Q$ and b) and d)
correspond to different $Q$ and constant $L$. Phonon dispersion curves are
shown in the insets.}
\end{figure}

\begin{figure}
\caption{Reduced {\it Z(XY)Y} Raman spectra of the disordered {\it PST} at
320 K: a) -- obtained by use of eq. 15 and 16, b) -- experimentally obtained. In a)
dotted lines correspond to modes used to describe the spectrum and solid line
corresponds to the resulting calculated specrum.}
\end{figure}

\begin{figure}
\caption{a) -- simulated unpolarized spectra 
with the disordered {\it PST} spectra (673~K) taken as a basis,
and b) -- unpolarized Raman spectra of {\it BMT} (from [22]) }
\end{figure}

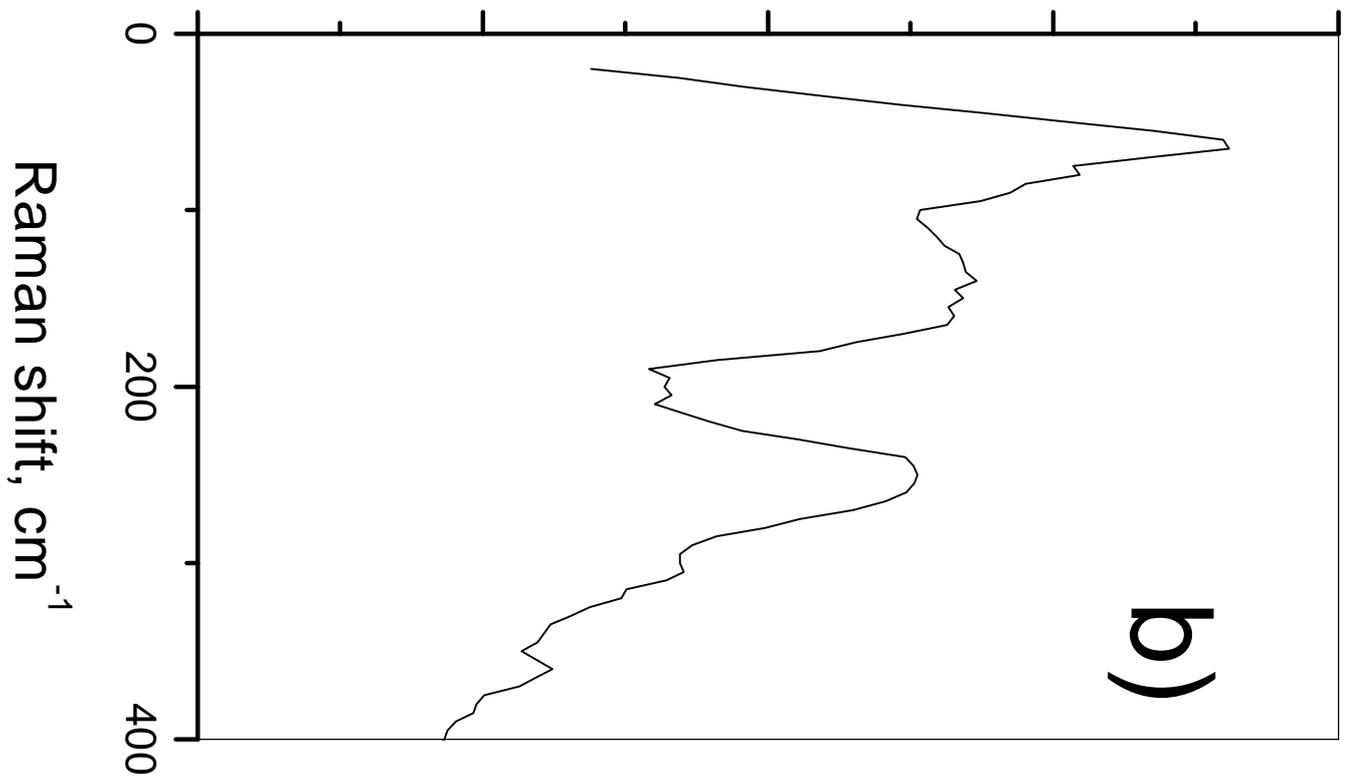
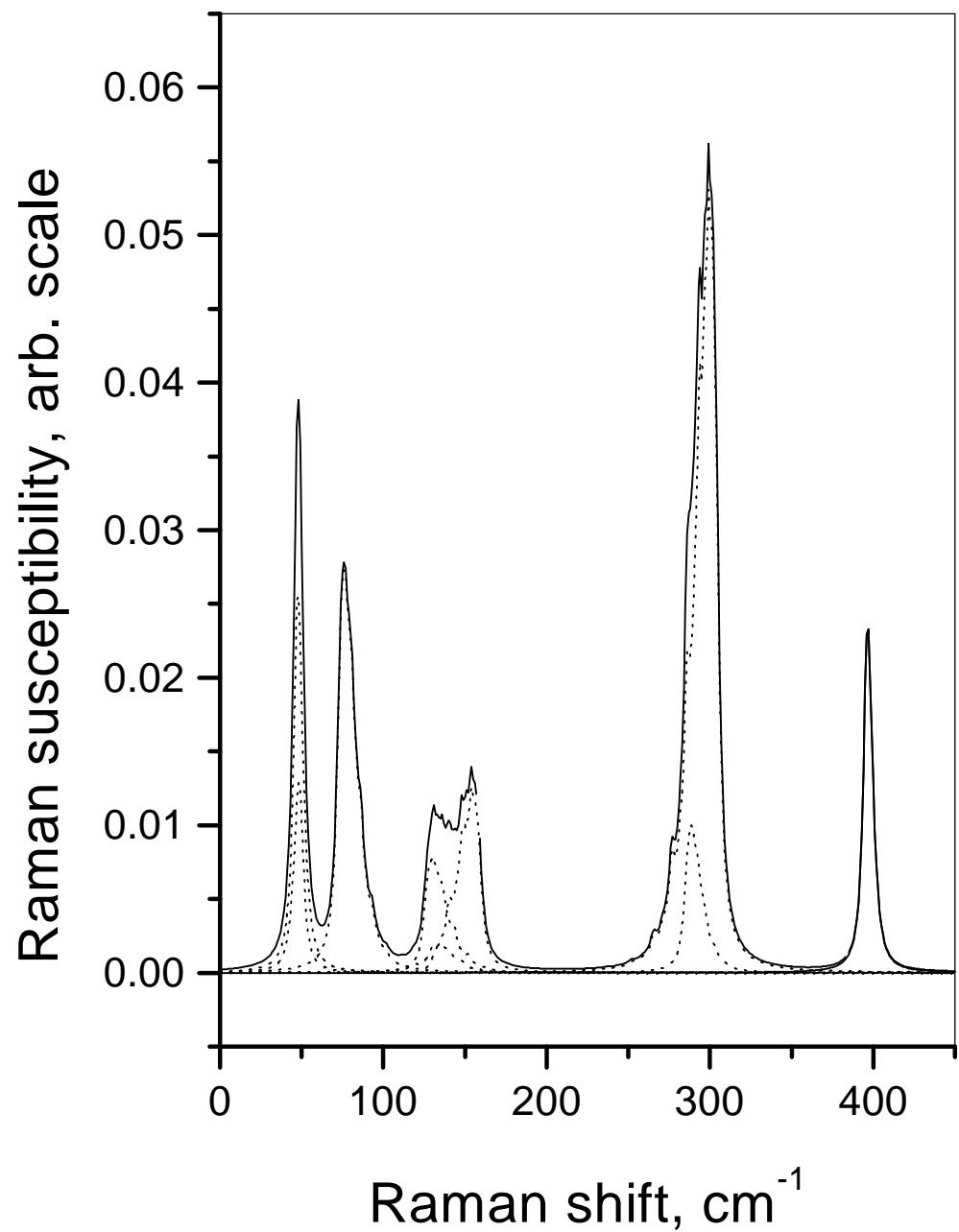
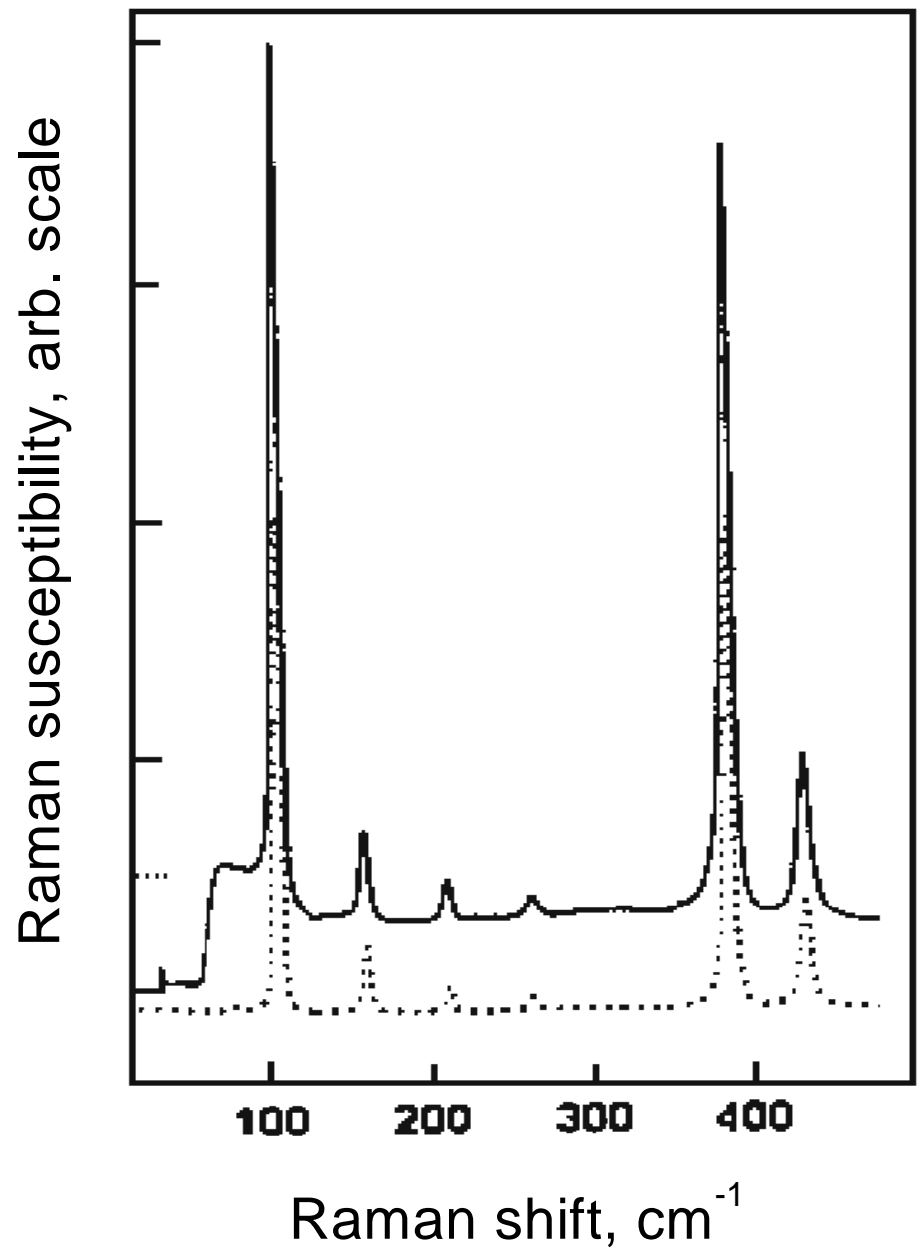
\end{document}